\documentstyle[aps,prl,epsf,floats]{revtex}

\begin{document}
\draft

\newcommand{\lsix} {La$_{1.94}$Sr$_{0.06}$CuO$_4$}
\newcommand{\lten} {La$_{1.90}$Sr$_{0.10}$CuO$_4$}
\newcommand{\lsco} {La$_{2-x}$Sr$_x$CuO$_4$}
\newcommand{\lnco} {La$_{1.4-x}$Nd$_{0.6}$Sr$_x$CuO$_4$}
\newcommand{\la} {$^{139}$La}
\newcommand{\cu} {$^{63}$Cu}
\newcommand{\cuo} {CuO$_2$}
\newcommand{\cutp} {Cu$^{2+}$}
\newcommand{\ybco} {YBa$_{2}$Cu$_3$O$_{6+x}$}
\newcommand{\ybcoca} {Y$_{1-x}$Ca$_x$Ba$_2$Cu$_3$O$_6$}
\newcommand{\etal} {{\it et al.}}
\newcommand{\ie} {{\it i.e.}}
\newcommand{\jpsj} {{J. Phys. Soc. Jpn.}}

\wideabs{

\title{Glassy Spin Freezing and NMR Wipeout Effect in the High-$T_c$
Superconductor La$_{1.90}$Sr$_{0.10}$CuO$_4$: What is the
Relationship With Stripes ?}

\author{M.-H. Julien$^*$}
\address{Dipartimento di Fisica "A. Volta" e Unit\'a INFM di Pavia,
Via Bassi 6, 27100 Pavia, Italy}
\address{Department of Physics and Astronomy, Ames Laboratory,
Iowa State University, Ames, Iowa 50011}
\address{Laboratoire de Spectrom\'etrie Physique, Universit\'e J.
Fourier, BP 87, 38402, Saint Martin d'H\'eres, France}

\author{A. Campana, A. Rigamonti, P. Carretta}
\address{Dipartimento di Fisica "A. Volta" e Unit\'a INFM di Pavia,
Via Bassi 6, 27100 Pavia, Italy}

\author{F. Borsa}
\address{Dipartimento di Fisica "A. Volta" e Unit\'a INFM di Pavia,
Via Bassi 6, 27100 Pavia, Italy}
\address{Department of Physics and Astronomy, Ames Laboratory,
Iowa State University, Ames, Iowa 50011}

\author{P. Kuhns, A.P. Reyes, W.G. Moulton}
\address{National High Magnetic Field Laboratory, 1800 E. Paul Dirac Dr., Tallahasse, FL 32310}

\author{M. Horvati\'c}
\address{Grenoble High Magnetic Field Laboratory, 38042 Grenoble
Cedex 9, France}

\author{C. Berthier}
\address{Laboratoire de Spectrom\'etrie Physique, Universit\'e J.
Fourier, BP 87, 38402, Saint Martin d'H\'eres, France}
\address{Grenoble High Magnetic Field Laboratory, 38042 Grenoble
Cedex 9, France}

\author{A. Vietkin and A. Revcolevschi}
\address{Laboratoire de Physico Chimie de l'Etat Solide, Universit\'e Paris-Sud,
91405 Orsay Cedex, France}

\date{\today}
\maketitle

\begin{abstract}

We report on \la~and \cu~NMR/NQR measurements in the high-$T_c$
superconductor La$_{1.90}$Sr$_{0.10}$CuO$_4$ with $T_c=26.5$~K.
Spin fluctuations probed by \la~spin-lattice relaxation ($T_1$),
continuously slow down on cooling through $T_c$. We argue that
spin-freezing and superconductivity are bulk effects in this
sample. Thus, both phenomena have to coexist microscopically. The
distribution of \la~$T_1$ values at low temperature reveals a wide
spread of spin fluctuation frequencies in CuO$_2$ planes. A simple
estimate shows that \cu~nuclei at sites where electronic
fluctuations are the slowest are not observable because of too
short relaxation times ({\it wipeout} effect). This means that the
\cu~NQR wipeout, observed in this sample, can be explained
primarily by slow magnetic, rather than charge, fluctuations. This
result does not rule out the connection between wipeout effect and
charge stripe order (Hunt \etal, (Phys. Rev. Lett. {\bf 82}, 4300
(1999)), but it indicates that the relationship between both
phenomena is not straightforward. We argue that the wipeout
fraction cannot define a proper order parameter for a stripe
phase, and cannot be used {\it alone} as a criterion for its
existence.

\end{abstract}


}

\section{Introduction}

Although \lsco~is probably the best studied high-$T_c$
superconductor (HTSC)~\cite{Johnston96,Kastner98} it still
continues to reveal new phenomena that were not observed or were
overlooked in the past. This is even true for microscopic and
powerful probes such as nuclear magnetic resonance (NMR), nuclear
quadrupole resonance (NQR) or muon spin rotation ($\mu$SR), which
have already made major contributions to our understanding of
these compounds~(for reviews,
see~\cite{Berthier96,Rigamonti98,Asayama98}). Two important
features have been highlighted recently:

(i) The coexistence of superconductivity with a frozen magnetic
state called a "cluster spin-glass" at concentrations
$0.06\lesssim x \lesssim 0.10$, an early result which was
confirmed by recent studies~(Refs.
\cite{Niedermayer98,Julien99,Panagopoulos00} and \cite{Julien99b}
for a concise review). In principle, the coexistence of
superconductivity with frozen spins or localized charges, is
rather hard to conceive. Indeed, in the cuprates, a link between
superconductivity and the characteristic frequency of spin
fluctuations~\cite{Berthier96,Asayama98,Julien96} was inferred
from NMR data at relatively high temperature, $T\gg T_c$. So, it
appears now equally important to characterize the evolution of
magnetic fluctuations close to and below $T_c$, in {\it e.g.}
\lsco.

(ii) A strong "wipeout effect" for the \cu~NQR signal at
concentrations $x<0.12$: the number of \cu~nuclei contributing to
the signal decreases on cooling (even above $T_c$) and the signal
completely disappears at low temperature~($T$)~\cite{Hunt99}.
Interestingly, a very similar phenomenology, with spin freezing
and Cu NQR/NMR wipeout, is observed
\cite{Hunt99,Singer99,Suh00,Curro00,Matsumura00} in cuprate
materials where doped holes have been shown to order in linear
single rows, known as charge {\it stripes}~\cite{Tranquada98}. In
fact, it was even discovered that the wipeout fraction (\ie~the
fraction of unobserved Cu nuclei) has the same $T$ dependence as
the "stripe order parameter" (actually the intensity of
superlattice peaks in neutron and X-ray scattering) in these
materials~\cite{Hunt99,Singer99}. As explained by Hunt \etal~the
wipeout effect can be caused by very slow (in the MHz range) {\it
charge and/or spin} fluctuations~\cite{Hunt99}. However, the
identification of the wipeout fraction as a stripe order
parameter~\cite{Hunt99} has led to the belief that the effect was
predominantly caused by the charges \cite{Service99} and that
stripe order in \lsco~was characterized by ultra slow charge
motion~\cite{AbuShiekah99}. This idea was recently challenged by
Curro \etal~who attribute the wipeout effect to slow spin
fluctuations exclusively and refute any relationship to the stripe
order parameter \cite{Curro00}. Since the discovery of Hunt
\etal~\cite{Hunt99} has raised the hope of having a new tool to
detect charge stripes (which have been so far elusive in most
materials), it is clearly important to better understand the
origin of the wipeout, and decide if this effect can be considered
as a criterion for stripe order.

Here, we report on \la~NMR/NQR and \cu~NQR measurements in \lten.
The main conclusions of this work are: (1)~bulk superconductivity
coexists with frozen magnetic moments throughout the sample.
(2)~The slow and inhomogeneous spin dynamics characterizing the
freezing process provide the most likely explanation for the Cu
NQR wipeout effect.

The paper is organized as follows: first, the main magnetic
properties of \lten~at low $T$ are recalled in Section II. Section
III gives a basic NMR background, focused on spin-lattice
relaxation and wipeout effects. Experimental details, including
discussion of the NMR lineshape, are described in Section IV, with
a brief account of magnetization measurements which indicate bulk
superconductivity in the sample. Section V is devoted to the
NMR/NQR results and to their analysis. The results are summarized
in Section VI, together with a discussion in a more general
perspective.

\section{Context of the work}

We precise the context of the experiment by summarizing some
magnetic properties of \lsco~with $x\simeq0.1$, focusing on
relatively low temperatures, $T\lesssim100$~K. Hunt \etal~have
reported a loss of Cu NQR signal, below $\sim$70~K for $x=0.09$
and below $\sim$50~K for $x=0.115$, in ceramic samples
\cite{Hunt99}. A similar wipeout effect may thus be anticipated
below $\sim$60~K for $x=0.10$.

Incommensurate elastic peaks are found in neutron scattering below
$\simeq$15~K~\cite{Kimura99}, \ie~in the superconducting state
($T_c\sim30$~K). By analogy with the results of Tranquada and
co-workers in La$_{1.48}$Nd$_{0.4}$Sr$_{0.12}$CuO$_4$
\cite{Tranquada95}, this modulated AF order is suspected to result
from the ordering of spin domains between antiphase walls formed
by charge stripes. However, the corresponding charge order peaks
have not been observed. On the other hand, a much lower
spin-ordering temperature of 1.2~K is reported from a $\mu$SR
study~\cite{Niedermayer98}. The difference between neutron and
$\mu$SR results is presumably ascribed to the glassy nature of
this ordering: spin fluctuations continuously slow down over a
wide $T$ range, so a dynamic measurement probes a
frequency-dependent ordering temperature. Actually, the existence
of frozen spins at $x=0.10$ was already inferred by Ohsugi
\etal~from the broadening of the NQR line at
1.4~K~\cite{Ohsugi94}. However, it was not clear if all or part of
the sample was magnetic. Slowing down of spin fluctuations is also
visible in EPR measurements \cite{Kataev93,Kochelaev97}.

In conclusion, \lten~lies at an interesting position in the phase
diagram of \lsco: while being close to the $x\sim0.12$ composition
where magnetic order is quite
strong~\cite{Kimura99,Torikai90,Kumagai94}, it shows magnetic
order only at quite low $T$ and has about two-thirds of the
highest $T_c$ achievable in this system (at ambient pressure). It
also shows a wide $T$ range of Cu NQR wipeout effect.

\section{NMR background}

\la~and \cu~are complementary NMR and NQR probes. \la~nuclei are
coupled to the magnetic moments of Cu$^{2+}$ electrons through the
hyperfine field which results from both a transferred contact
interaction (via orbital overlap with the apical oxygen) and a
direct dipolar interaction. The hyperfine field is estimated to be
$\simeq1-2$ kOe/$\mu_B$ \cite{Nishihara87}. In contrast, \cu~and
$^{65}$Cu nuclei experience a much larger coupling to both the
on-site Cu$^{2+}$ spin (anisotropic hyperfine field $A_c\simeq
-134$~kOe/$\mu_B$, $A_{ab}\simeq 2$~kOe/$\mu_B$~\cite{Walstedt95})
and to the four Cu$ ^{2+}$ first neighbors (isotropic transferred
coupling $B\simeq 35$~kOe/$\mu_B$~\cite{Walstedt95}).

\subsection{Nuclear relaxation}

Nuclear spin-lattice relaxation occurs through temporal
fluctuations of the local magnetic field (magnetic relaxation)
and/or of the electric field gradient (quadrupolar relaxation).
For magnetic relaxation, the spin-lattice relaxation rate 1/$T_1$
of a given nucleus is proportional to the {\it square} of its
gyromagnetic ratio $\gamma_n$ and to the {\it square} of the
transverse components ($h_x$, $h_y$) of the local field (the $z$
quantization axis is the direction of the external magnetic field
$H$) :

\begin{equation}
\frac{1}{T_1} = \frac{\gamma_n^2}{2} \int_{-\infty}^{+\infty}
\overline{\langle h_+(t)h_-(0)\rangle}
 \exp(i\omega_nt) dt  , \label{T1}
\end{equation}
where the horizontal bar denotes the ensemble average and
$\omega_n$ is the nuclear Larmor frequency.

With $^{63}h_\perp^2/^{139}h_\perp^2 \gtrsim5000$ for $H\|c$ and
with $(^{63}\gamma/^{139}\gamma)^2\simeq 3.5$, one finds
immediately that fluctuations of Cu$^{2+}$ spins will lead to
$T_1$ values for \cu~which are shorter by about four orders of
magnitude compared to the values for \la~(the exact number depends
on the wave-vector dependence of spin fluctuations, which are
ignored in the above estimate).

For fluctuations of the local field of the form $\overline{\langle
h_+(t)h_-(0)\rangle}=\langle h_\perp^2\rangle\exp(-2t/\tau_c)$,
where $\tau_c$ is called the correlation time, a standard
expression for $T_1$ can be derived \cite{Slichter91}:
\begin{equation}
\frac{1}{T_1} = \gamma_n^2 \langle h_\perp^2\rangle
\frac{2\tau_c}{1+\omega_n^2\tau_c^2} , \label{BPP}
\end{equation}

Slowing down of magnetic fluctuations means that $\tau_c$
increases on cooling  ($\tau_c^{-1} \gg\omega_n$), and this leads
to an increase of $T_1^{-1}$, which eventually reaches a maximum
when $\tau_c^{-1}=\omega_n$ (Eq.~\ref{BPP}). The existence of a
maximum in $T_1^{-1}$ as a function of $\tau_c$ is a rather
general feature (more general than the particular form of $T_1$
assumed above), which also holds when the system cannot be
described by a single value of $\tau_c$, but rather by a
distribution of correlation times. The maximum of $1/T_1$ {\it
vs.} $T$ becomes broader in this case. The temperature at which
$1/T_1$ reaches a maximum defines the freezing temperature $T_g$
at the NMR time scale. When slowing down occurs over a rather wide
temperature range, the value of $T_g$ determined by another
experimental technique, with a different time scale, may differ
significantly from the NMR value. With typically
$\omega_n^{-1}\sim 10^{-8}$~s, \la~NMR is a relatively slow probe,
with a timescale comparable to $\mu$SR.

The spin-spin relaxation time $T_2$ defines the characteristic
time decay of the echo height in a spin-echo sequence. So, $T_2$
determines the time available for recording the NMR signal. In
most solids, $T_2$ is determined by nuclear dipole-dipole
interaction. In the cuprates, $T_2$ of $^{63,65}$Cu nuclei is
dominated by two stronger processes: $T_{2G}$, which comes from
indirect exchange between Cu nuclei via the non-local electronic
susceptibility \cite{Pennington91} and {\it $T_{2R}$, the Redfield
contribution, which is a function of
$T_1$}~\cite{Slichter91,remRedfield}. Since both processes are
proportional to squares of hyperfine fields, \la~nuclei have a
much longer $T_2$ ($\sim$ms) than $^{63,65}$Cu nuclei ($\sim 1-50
\mu$s).

\subsection{Wipeout effects in NMR/NQR}

The NMR/NQR signal is directly proportional to the population
difference between adjacent nuclear levels. As such, it is
proportional to $1/T$ and to the number of nuclei in the sample.
In practice, since the observation occurs at a finite time after a
radio-frequency pulse, the measured signal is reduced from its
maximum possible value because of the $T_2$ process. The decrease
is typically of Lorentzian or Gaussian type. So, in order to check
if all nuclei are observed as a function of temperature, the
signal should be renormalized by a factor $T$ and then corrected
for the $T_2$ effect by extrapolating its magnitude at time zero.

The term {\it wipeout} effect was introduced in order to describe
the loss of NMR signal due to non-magnetic impurity doping in
metals~\cite{Bloembergen53}. The decrease of the NMR signal
occurred because of the large spread of resonance frequencies out
of a given spectral window. Similar loss of NMR signal has been
well-documented for localized magnetic moments in metals (RKKY
oscillations)~\cite{Winter71}. A transition to a magnetically
ordered phase may also lead to an apparent loss of signal, because
the internal field causes a shift of the resonance positions, with
possibly sizeable broadening.

On the other hand, a loss of signal may be produced by a dramatic
shortening of $T_2$ (or of $T_1$, through the $T_{2R}$ term). The
correct signal cannot be obtained by extrapolating the measured
signal at time zero, when relaxation times for part of the nuclei
become shorter than the "dead time" of the spectrometer, \ie~some
nuclei have relaxed so fast that the signal coming from them
cannot be digitized. If this occurs for all nuclei in the sample,
the signal is completely lost. As seen above, very short
relaxation times occur if the spectral density of electronic
fluctuations is large at the Larmor frequency.

Finally, when the Larmor frequency  is directly defined by the
value of the hyperfine electric or magnetic coupling (as in NQR or
zero field NMR), strong fluctuations of these couplings at the
observation time scale will also cause a loss of the resonant
signal. Because of the slow fluctuations involved, very short
nuclear relaxation times are likely to occur in such cases. Thus,
the various contributions to the wipeout effect may not be
distinguishable.

In summary, a loss of NMR signal can result from static effects
(modification of the lineshape), from dynamical effects or from
both. These effects can occur homogeneously or inhomogeneously in
the sample. Inhomogeneity frequently leads to a wipeout effect
which is only partial, in which case careful measurements are
required in order to realize that part of the signal has been
lost. Not surprisingly, wipeout effects are observed in canonical
spin-glasses close to the glass
transition~\cite{Levitt77,MacLaughlin77,Chen83}.

\section{Experimental details}

\subsection{NMR measurements}

\begin{figure}[t!]
\begin{center}
\epsfxsize=90mm
 $$\epsffile{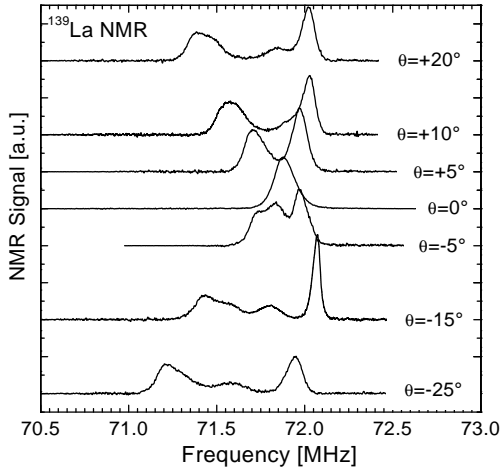}$$
% \vspace{-1cm}
\caption{\la~NMR spectra in a field $H$=12 Tesla at $T$=~57 ~K.
$\theta$ is the angle between $H$ and the $c$-axis. The line
splitting is related to different directions of the main axis of
the electric field gradient (see text).} \label{angle}
\end{center}
\end{figure}

Most of the experiments were performed on a crystal ($m=113$~mg)
grown by the traveling solvent floating zone method~\cite{Revco}.
It is a piece of the large crystal used by Petit \etal, for
neutron scattering measurements~\cite{Petit98}. In the course of
the NMR experiments, it was found that the sample was not a true
single crystal, as a part of it had a different orientation from
the rest. While this does not affect our analysis of the NMR/NQR
signal intensity (because all frequencies were integrated), it
affects the lineshapes. This sample was then cut into two equal
pieces, one of which was confirmed to be a single crystal of very
high quality from both neutron and X-ray diffraction. We then
performed new \la~NMR lineshape measurements as well as \la~$T_1$
and $T_2$ measurements on a well-isolated line in this single
crystal ($\theta=-15^\circ$ in Fig.~\ref{angle}). The recovery
laws were strictly identical to those in the original sample.
Thus, the distribution of $T_1$ values, which will be discussed
below, is intrinsic. Within experimental accuracy, $T_1$ and $T_2$
are the same on the different lines.
%%%%%%%%%%%%%%%%%%%%%
%\vspace{-9mm}
%%%%%%%%%%%%%%%%%%%%%%%%%%%%%%%%%%%%%%%%%%%%%%
%%%%%%%%%%%%%%%%%%%%%%
%\vspace{-9mm}
\begin{figure}[h!]
\begin{center}
\epsfxsize=80mm
 $$\epsffile{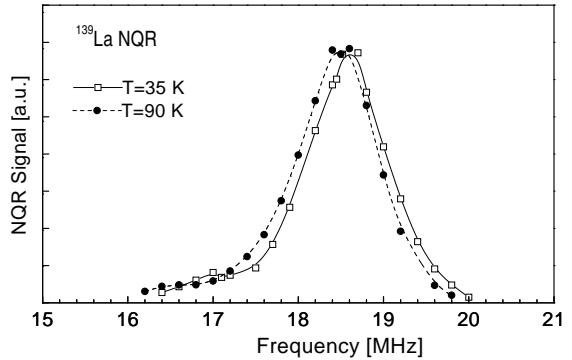}$$
\vspace{-0.8cm} \caption{High frequency (3$\nu_Q$) transition of
the \la~NQR spectra at 90~K and 35~K.} \label{lanqrsp}
\end{center}
\end{figure}
%%%%%%%%%%%%%%%%%%%%

NMR spectra were obtained from a single Fourier transform of half
of the spin echo signal when the line was sufficiently narrow. For
lines broader than the frequency window of the excitation, spectra
were obtained by summing Fourier transforms at equally spaced
frequencies or by recording the spin-echo integral/amplitude at
different frequencies. \cu~and \la~NQR spectra were recorded with
the latter method.

The \la~NMR spectrum with $H\|c$ in the single crystal shows a
single peak which splits into four peaks as the magnetic field is
tilted away from the $c$-axis (one of the lines is not
well-defined, but only appears as a shoulder; see
Fig.~\ref{angle}). The angular dependence of this spectrum
(Fig.~\ref{angle}) and the fact that the \la~NQR spectrum does not
show well separated lines but a single peak (Fig. \ref{lanqrsp})
indicate that the splitting is due to different values of the
angle between $H$ and the direction of the principal axis of the
electric field gradient tensor $V_{zz}$. Since these four angles
are roughly equal when $H\|c$ (Fig.~\ref{angle}), the effect
originates from different directions of $V_{zz}$ in the crystal,
with equal tilts from the $c$-axis. Unambiguous identification of
the different directions would require a full angle-dependence
study within two perpendicular rotation planes. Because of sample
geometry, this could not be performed here. The various directions
of $V_{zz}$ may correspond to different tilt directions of CuO$_6$
octahedra. The $T$-dependence of the shifts and linewidth show
that the tilt angles increase smoothly on cooling from~$\simeq
340$~K down to~$\simeq$100~K, with a saturation at lower $T$.

These results motivated us to reexamine \la~NMR spectra in the
single crystal of \lsix~that we used for a previous study
\cite{Julien99}. The new measurements, at higher magnetic fields
and with improved experimental conditions, revealed that the
spectrum is composed of at least three lines, whose separation is
predominantly of quadrupolar origin, as in \lten. Thus, this
finding invalidates the hypothesis in Ref.~\cite{Julien99} of only
two lines split by a purely magnetic effect.

Measuring the magnitude of the signal intensity requires care. In
order to ensure that both the radio-frequency excitation and the
detection of the signal remain identical at all temperatures,
experimental conditions such as the 50~$\Omega$ matching and the
$Q$ factor of the resonant circuit should be carefully controlled.
Here, this was made easier by the fact that the capacitors for
impedance matching and frequency tuning were outside the probe, at
the constant room temperature.

Because of flux expulsion in the superconducting state, the NMR
signal in a single crystal is reduced drastically below $T_c$. In
order to study the spin dynamics down to low temperature, we have
applied high magnetic fields up to $H=23.2$~T. Such a field is
expected to reduce $T_c$ down to a few Kelvin, although the exact
$T_c$ is not known here, and in any case hard to define given the
broadening of the transition under field. The experiments at
23.2~T were carried out in a high homogeneity resistive magnet of
the National High Magnetic Field Laboratory in Tallahassee, FL.
Other measurements up to 15~T were carried out in superconducting
coils.

%%%%%%%%%%%%%%%%%%%%%
%\vspace{-9mm}
\begin{figure}[t!]
\begin{center}
\epsfxsize=75mm
 $$\epsffile{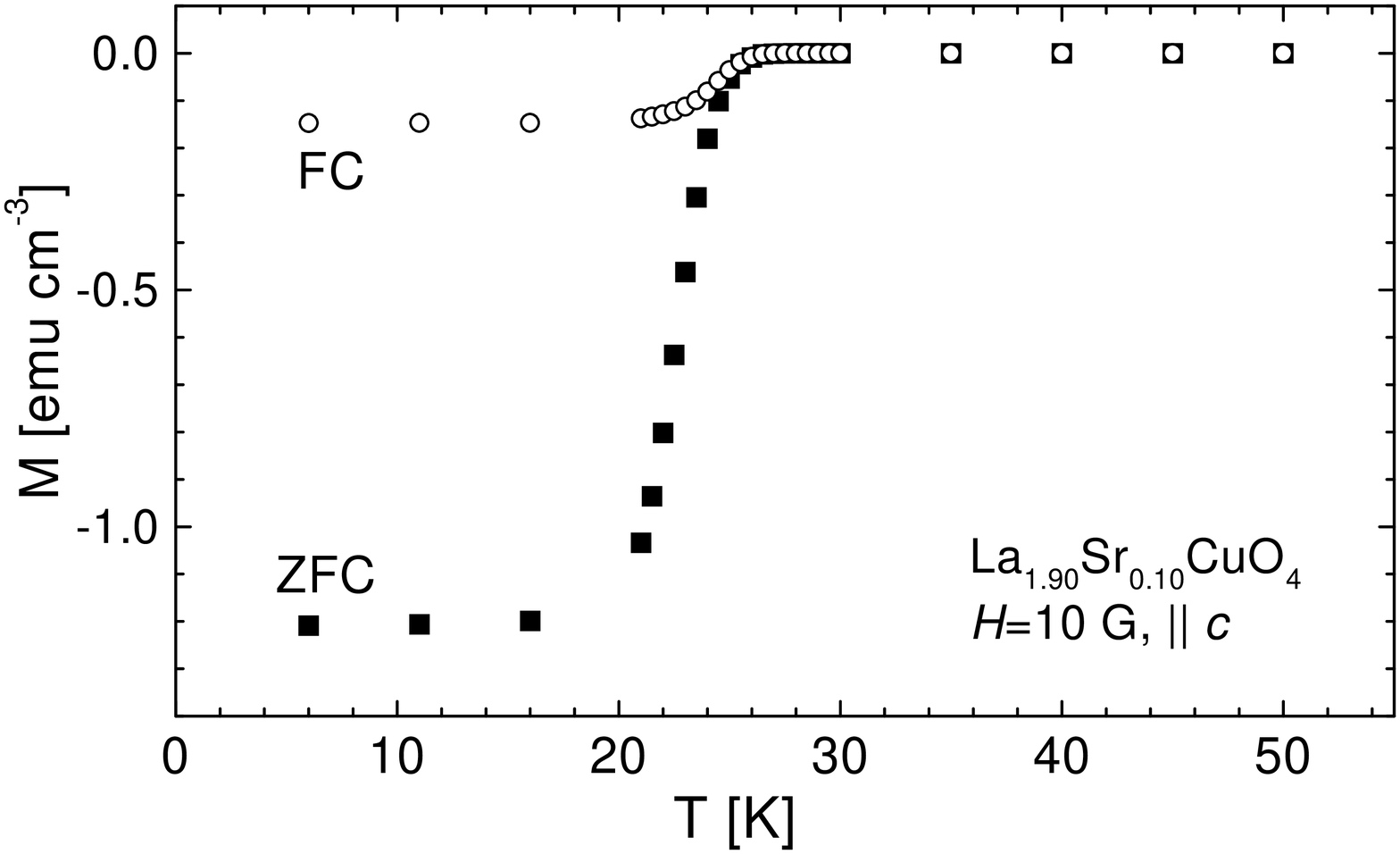}$$
 \vspace{-1cm}
\caption{Field cooled (FC) and zero-field cooled (ZFC)
magnetization of the \lten~single crystal (from Huh
\etal~\protect\cite{Huh00}).} \label{tc}
\end{center}
\end{figure}
%%%%%%%%%%%%%%%%%%%%%%%%%%%%%%%%%%%%%%%%%%%%%%

\subsection{Superconducting properties}

Magnetization measurements were performed on the polycrystalline
sample and on the single crystal, with almost identical results.
In Fig.~\ref{tc} results for the single crystal are
reported~\cite{Huh00}. The sample quality can be checked from the
narrow width ($\simeq 5$~K) of the superconducting transition,
which has an onset at $T_c=26.5$~K. It is particularly difficult
to establish bulk superconductivity from single crystal
measurements in fields much higher than a few
G~\cite{Nagano93,Moodenbaugh97}. Nevertheless, a study of the high
field reversible magnetization, similar to that in
Ref.~\cite{Ostenson97}, concludes that our sample is a bulk
superconductor~\cite{Huh00}.
%%%%%%%%%%%%%%%%%%%%%%%%%%%%%%
\begin{figure}[t!]
\begin{center}
\epsfxsize=125mm
 $$\epsffile{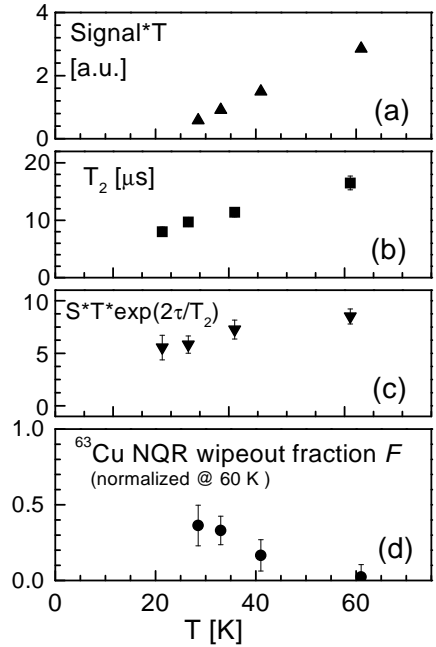}$$
\caption{(a) $T$-normalized signal intensity from the \cu~NQR
spectra, all taken in the very same experimental conditions. (b)
$T$ dependence of $T_2$ for \cu. (c) $T$- and $T_2$-normalized
signal intensity. (d) Wipeout fraction (see text), which is a
measure of the unobserved signal.}\label{wipeout}
\end{center}
\end{figure}
%%%%%%%%%%%%%%%%%%%%%%%%%%%%%%

\section{NQR and NMR results}

\subsection{\cu~NQR spectra: wipeout effect}

As a preliminary step of this study, we intended to check the
presence of a wipeout effect, as previously reported by Hunt
\etal~\cite{Hunt99}. The \cu~signal was recorded in an NQR
experiment, \ie~in zero external magnetic field. Because of the
excessive loss of signal in the superconducting state, our data
are limited to a narrow $T$ range above $T_c=26.5$~K. $T_2$ was
measured at different frequencies on the NQR spectrum, and was
found to shorten on decreasing frequency, in agreement with
Ref.~\cite{Hunt99}. The frequency dependence is however relatively
weak so that it is sufficient to correct the frequency-integrated
signal from the value of $T_2$ at the center of the line
only~\cite{Singer99}. $T_2$ was estimated from a single
exponential fit [$s(t)\propto \exp(-\frac{2\tau}{T_2}$)] of the
echo decay (the accuracy of the data did not allow to distinguish
between Lorentzian and Gaussian forms of the echo decay).

Fig.~\ref{wipeout} shows the results for the signal intensity,
$T_2$ and the wipeout fraction $F(T)=[{\rm s^*(60 K)-s^*}(T)]/{\rm
s^*}(60 K)$ ($\rm s^*$ is the signal corrected for $T_2$ and
temperature, and integrated over frequencies). There is clearly a
wipeout effect, which basically agrees with the results of Hunt
\etal~\cite{Hunt99} (our values appear somewhat lower, presumably
because of the normalization to the intensity at 60 K, our highest
temperature data point; Hunt \etal~find 20\% of wipeout for
$x=0.09$, at this temperature).
%%%%%%%%%%%%%%%%%%%%%%
%\vspace{-9mm}
\begin{figure}[t!]
\begin{center}
\epsfxsize=90mm
 $$\epsffile{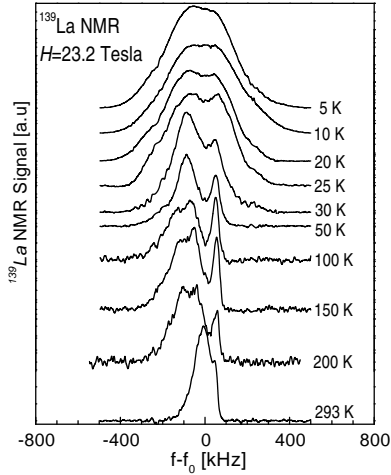}$$
\caption{\la~NMR spectra (central transition) in a field of
$H$=23.2~T ($f_0$=139.528~MHz). The splitting is due to the
combined effect of sample mosaicity and small misalignment (see
Section "experimental details"). Note the broadening below
$\sim$50~K.} \label{lanmr23t}
\end{center}
\end{figure}
%%%%%%%%%%%%%%%%%%%%

\subsection{\la~NMR spectra: low $T$ magnetic broadening}

As shown in Fig.~\ref{lanmr23t}, the \la~central line in a field
of 23.2~T broadens on cooling below about 50~K, with a saturation
of the width below $\simeq$10~K. The same broadening is seen at
15~T \cite{rembroad} (not shown), but not at 9.4~T above 15~K
(Fig.~\ref{nowipeout}). This strongly suggests that the broadening
is of magnetic origin, as previously observed in
\lsix~\cite{Julien99}. The broadening is not seen at 9.4~T,
because the linewidth is dominated by the large quadrupolar
broadening (which varies as 1/$H_0$). The \la~NMR broadening is
not an artifact related to signal loss at the center of the
spectrum (see next subsection), and is not of dynamic origin
($^{139}T_2$ is much longer than the inverse linewidth in the
single crystal), at least above $\sim$10~K. So, the broadening
indicates a spread of local fields along the $z$ direction. This
is a purely paramagnetic effect since $T_1$ data show no sign for
frozen moments in the range 10-60~K, but a smooth evolution toward
freezing at lower $T$ (see Fig.~\ref{T1vsT} and discussion below).
Clearly, the broadening cannot be caused by frozen magnetic
regions at so high temperatures. This effect is somehow similar to
the one observed in Zn-doped materials~\cite{JulienZn} and is not
necessarily seen in bulk magnetization measurements, which sums
the contributions from all staggered moments.

Note that the above discussion makes clear that {\it the
broadening of the NMR line cannot be taken as evidence for
magnetic order}. This erroneous criterion has been sometimes used
to define a N\'eel temperature in \lsco~with
$x\simeq0.12$~\cite{Goto94,Goto97,Suzuki98}.

\subsection{\la~NMR spectra : absence of wipeout effect}
%%%%%%%%%%%%%%%%%%%%%%%%%%%%%%%%%%%%%%%%%%%%%%
\begin{figure}[t!]
\begin{center}
\epsfxsize=80mm
 $$\epsffile{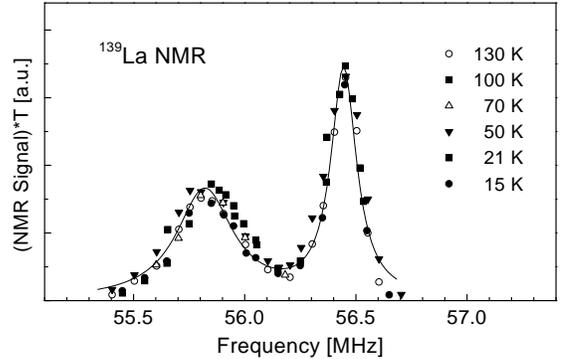}$$
\caption{$T$-normalized NMR signal for the \la~NMR central line.
The solid line is a guide to the eye. Except for minor changes,
the data collapse on a single curve. There is no wipeout effect in
this temperature range.} \label{nowipeout}
\end{center}
\end{figure}
%%%%%%%%%%%%%%%%%%%%%%
The $T$ dependence of the \la~intensity was measured for the NMR
central line ($m_I=+1/2 \leftrightarrow$ -1/2 transition) in a
field of 9.4~T. $T_2$ was observed to shorten on cooling. However,
{\it in the range 100-15~K}, $^{139}T_2$ of the order of ms
remains much longer than the delay between $\pi/2$ and $\pi$ NMR
pulses ($\tau\simeq 30 \mu$s), so the $T_2$ correction is
essentially negligible for~\la.

As shown in Fig.~\ref{nowipeout}, the NMR signal (multiplied by a
factor $T$) is independent of $T$ from 130~K down to 15~K. There
is no loss of \la~NMR signal on cooling in the range 70~K-15~K,
where the Cu NQR spectrum is wiped out. This contrasting behavior
is not {\it a priori} unexpected since \la~lies out of the CuO$_2$
planes. Hence, spin and charge fluctuations in these planes
produce hyperfine field and electric field gradient fluctuations
which are considerably reduced at the La site. Still, the absence
of \la~NMR wipeout down to 15~K makes $^{139}T_1$ measurements
particularly interesting, since the whole sample is probed,
including those parts where the Cu NQR signal has disappeared.

%%%%%%%%%%%%%%%%%%%%%%
\begin{figure}[t!]
%\vspace{-25mm}
\begin{center}
\epsfxsize=80mm  $$\epsffile{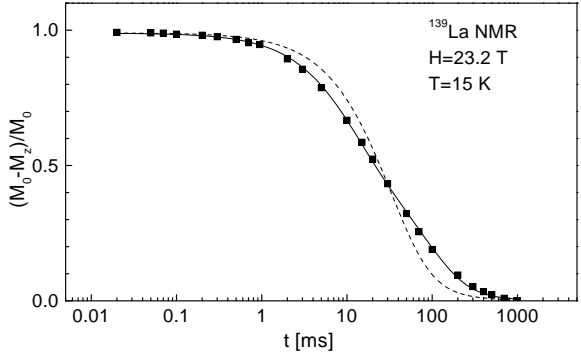}$$% \vspace{-5.5cm}
\caption{Time dependence of the \la~NMR signal $M_z$ after
saturation of the central line at $T=19$~K, in a field of
23.2~Tesla ($M_0=M_z(t=\infty)$). The dashed and solid lines are
fits with one and two components respectively, with each component
following the theoretical recovery law for magnetic relaxation
(see text).} \label{distribT1}
\end{center}
\end{figure}
%%%%%%%%%%%%%%%%%%%%%%

\subsection{\la~NMR relaxation evidence for a distribution of spin fluctuation frequencies}

A typical plot of the time dependence of the \la~longitudinal
magnetization after a comb of saturation pulses at $T=15$~K is
shown in Fig.~\ref{distribT1}. For a purely magnetic relaxation
mechanism, the theoretical expression for the recovery of the
magnetization, after fast irradiation of the central NMR
transition of nuclear spins $I=\frac{7}{2}$, is \cite{Narath67} :
$[M_0-M_z]/M_0=[8575\exp(-28t/T_1)+
2475\exp(-15t/T_1)+819\exp(-6t/T_1)+143\exp(-t/T_1)]/12012$.
However, the data points cannot be fitted to this expression
(Fig.~\ref{distribT1}). This means that the recovery is modified
by a distribution of $T_1$ values or by nuclear transitions
produced by electric field gradient fluctuations. At low
temperatures, $T_1$ is entirely magnetic (see below the large
enhancement of $T_1^{-1}$ due to the spin-freezing). So, the
deviation of the recovery from the ideal behavior is due to a
distribution of $T_1$ values in that case. Such a distribution was
previously found to characterize the magnetic freezing in
\lsco~($x>0.02$)~\cite{Julien99,Cho92} and in
La$_{1.65}$Eu$_{0.2}$Sr$_{0.15}$CuO$_4$~\cite{Curro00}. Since the
shape of the \la~ recovery smoothly depends on $T$ in the range
5~K-40~K, it is very reasonable to assume that there is a
distribution of $T_1$ values in all this $T$
range~\cite{rem-quad}.

%%%%%%%%%%%%%%%%%%%%%%%%%%%%%%%%%%%%%%%%%%%%%%%%%%%
%\vspace{-5mm}
\begin{figure}[t!]
\begin{center}
\epsfxsize=80mm
 $$\epsffile{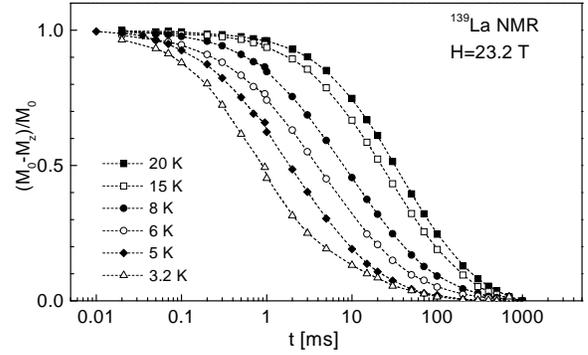}$$
\caption{Time dependence of the \la~NMR signal $M_z$ after
saturation of the central line in a field of 23.2~Tesla, at
different temperatures. The faster recovery of $M_z$ at low $T$ is
due to the slowing down of magnetic fluctuations.}
\label{recov23T}
\end{center}
\end{figure}
%%%%%%%%%%%%%%%%%%%%%%%%%%%%%%%%%%%%%%%%%%%%%%%%%%%

\subsection{Implications of \la~$T_1$ results
for the \cu~NQR/NMR wipeout effect}

Given the distribution of $T_1$ values, the data should be
characterized by the width and the central value of this
distribution. However, as it is usually observed in such cases,
the data may be reasonably well-fitted with only two contributions
(here of roughly equal weight), each of them following the
theoretical expression for $I=\frac{7}{2}$ (given above). We adopt
here this procedure for simplicity. From the data at 15~K, we
extract two characteristic relaxation times for \la~nuclei, which
have quite different values: $T_1^A=2200$~ms and $T_1^B=265$~ms.
With the help of Eq.~\ref{BPP}, one readily finds that magnetic
fluctuations responsible for a $T_1$ of 265~ms at La sites would
produce a $T_1\lesssim 15 \mu$s at \cu~sites and thus a
$T_2\lesssim 2 \mu$s~\cite{remRedfield}. The spin-echo signal from
nuclei with so short relaxation times can certainly not be
observed, while nuclei with relaxation times greater by an order
of magnitude should be observable. This implies that the \cu~NQR
spectrum will be partially wiped out because some nuclei have too
fast relaxation times to be observed, while the remaining nuclei
are still detected. It could be remarked that $^{139}T_1^{-1}$ is
weakly $T$ dependent down to 30-40~K, while Cu NQR wipeout starts
at least below 60~K. There is however no contradiction between
these two observations. Indeed, $T_1$ of \la~contains a background
of quadrupolar relaxation which likely masks the onset for the
enhancement of magnetic relaxation. Moreover, the quadrupolar
relaxation channel which is active for \la~nuclei is not
necessarily present at Cu sites. \la~and \cu~nuclei certainly have
a different ratio of magnetic to quadrupolar relaxation, and this
ratio is extremely difficult to determine
experimentally~\cite{Suter98}. Actually, $^{139}T_2$ shortens by a
factor of 1.3 between 60~K and 30~K.

Thus, we conclude that {\it the slow and spatially distributed
magnetic fluctuations are sufficient to explain the strong Cu
wipeout effect at low temperatures in \lten}. One cannot exclude
that slow charge fluctuations are present, but there is no
evidence for this.

\subsection{\la~$T_1$ evidence for spin-freezing}

%%%%%%%%%%%%%%%%%%%%%%%%%%%%%%%%%%%%%%%%%%%%%%

\begin{figure}[t!]
\begin{center}
\epsfxsize=80mm
 $$\epsffile{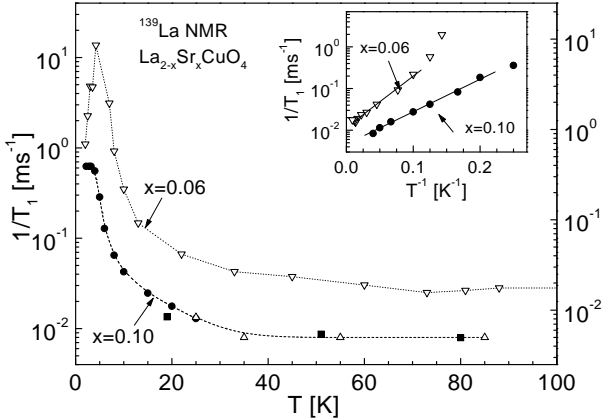}$$
\caption{T dependence of the \la~nuclear spin-lattice relaxation
rate $1/T_1$ \protect\cite{remT1value} for $x=0.10$ (this study)
and $x=0.06$ \protect\cite{Julien99}. The left scale is associated
with the NMR data for $x=0.10$: the filled circles are data at
23.2~T and the filled squares are data at 9.4~T. The right
vertical scale is associated with NQR data (up triangles for
$x$=0.10 and down triangles for $x=0.06$. Dotted and dashed lines
are guides to the eye. Inset: same data {\it vs.} 1/$T$;
continuous lines are fits explained in the text.} \label{T1vsT}
\end{center}
\end{figure}
%%%%%%%%%%%%%%%%%%%%%%

Fig.~\ref{recov23T} shows the recovery plots of the nuclear
magnetization after a sequence of saturation pulses, at selected
temperatures, in a field of 23.2~T. The overall trend clearly
shows that the recovery becomes faster on cooling, \ie~$T_1$
shortens. This behaviour holds down to about 4~K. Below 4~K, a
tail appears at long times, becoming as important as the fast
component at $T=1.65$~K, our lowest $T$ data point (not shown).
This long component in the relaxation does not seem to be linked
to superconductivity since it is also observed at 15~T in the same
$T$ range, and $T_c$ is higher at this field (at 15~T, a clear
change in the frequency tuning of the NMR probe at 6.5~K signals
the irreversibility line). More data are necessary in order to
understand if this feature, which might be caused by a
distribution of freezing temperatures, is intrinsic or related to
the sample purity. Because of this uncertainty and of the limited
experimental $T$ range, it is not possible to determine a precise
freezing temperature $T_g$ in this sample. Nevertheless, it is
clear that a slowing down phenomenon starts below $\sim30-40$~K,
as is shown in Fig.~\ref{T1vsT} by the $T$-dependence of
$T_1$~\cite{remT1value}. An important feature of the freezing
process is that it involves the vast majority of sites, if not
all, in the sample: there is no long tail in the recoveries
(excepted the one very close to $T_g$) which could be attributed
to non-freezing areas. In contrast, the recovery curves in
Fig.~\ref{recov23T} shift continuously toward short times on
cooling, without strong modification of their shape. Note that
this conclusion is not affected by a possible loss of \la~nuclei
below 15~K, since this would precisely originate from freezing
zones, while non-freezing ones have longer $T_1$.

\subsection{Remark on high magnetic fields}
It is important to realize that the spin freezing is not induced
by the magnetic field. In fact, (i) a similar enhancement of
$1/T_1$ was found in fields of 23.2, 15 and 9~T (not shown). (ii)
$\mu$SR measurements (in zero field) have already reported a bulk
spin-freezing at $T_g\simeq1.2$~K for
$x$=0.10~\cite{Niedermayer98} (see also~\cite{Ohsugi94}). Thus,
magnetic fields as high as 23~Tesla do not seem to modify spin
dynamics in \lten~for temperatures in the range 5-40~K (an
influence of the field on the freezing temperature $T_g$ cannot be
excluded).

It is interesting to consider these findings in comparison with
the insulating behavior of the $ab$-plane resistivity under strong
magnetic fields. Boebinger \etal~\cite{Boebinger96} find
$\rho_{ab}\propto \log 1/T$, which they consider as an indication
of non-metallic ground state in zero field when superconductivity
is destroyed, while Malinowski \etal~\cite{Malinowski97} suggest
that it is the field itself which induces localization of the
charges. Since spin dynamics are likely to be affected by charge
localization, our result that spin dynamics in the freezing regime
do not change appreciably with the field would rather tend to
support the former view.

\subsection{Comparison to other works}
%%%%%%%%%%%%%%%%%%%%%%%%%%%%%%%%%%%%%%%%%%%%%%%%%%%%%%%%%%%%%%%%%%%%
\begin{figure}[h!]
%\begin{center}
 \epsfxsize=110mm
 $$\epsffile{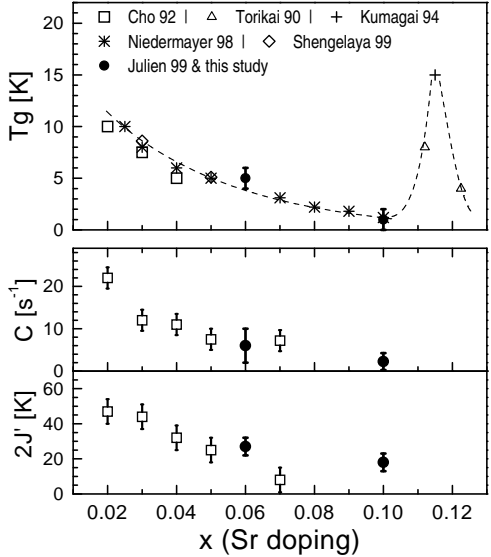}$$
\caption{Top panel: Magnetic transition temperature $T_g$ in
\lsco~($0.02<x<0.125$) from
$\mu$Sr~(\protect\cite{Niedermayer98,Torikai90,Kumagai94,Shengelaya99})
and NMR/NQR~(\protect\cite{Julien99,Cho92} and this work). $T_g$
data from magnetization measurements can be found in
\protect\cite{Filipkowski90,Chou95,Wakimoto99}. Middle and bottom
panels: Parameter $C$ and $J^\prime$, obtained from a fit
explained in the text, and compared to the data of Cho
\etal~(squares)~\protect\cite{Cho92}.} \label{Tg}
%\end{center}
\end{figure}
%%%%%%%%%%%%%%%%%%%%%%
The top panel of Fig.~\ref{Tg} shows a selection of spin-freezing
temperatures ($T_g$) determined from $\mu$Sr and NMR/NQR
experiments in
\lsco~(\cite{Niedermayer98,Torikai90,Kumagai94,Shengelaya99};
$T_g$ data from magnetization measurements can also be found
in~\cite{Filipkowski90,Chou95,Wakimoto99}). Given the
uncertainties discussed above, we plotted $T_g=1\pm1$~K for our
experiment in \lten, which is consistent with previous works, in
particular $T_g\simeq 1.2$~K from Niedermayer
\etal~\cite{Niedermayer98}.

Cho \etal~remarked that $T_1$ data for $T>2T_g$ can be fitted to
the following expression~\cite{Cho92}:
%%%%%%%%%%%%%%%%
\begin{equation}
\frac{1}{T_1} \propto c(x) \exp\left(\frac{2J^\prime(x)}{T}\right)
\label{eqncho}
\end{equation}
%%%%%%%%%%%%%%%
where $2J^\prime$ can be considered as a coupling constant. The
values of the parameters $c(x)$ and $J^\prime(x)$, obtained by
fitting our data for $x=0.10$ and $x=0.06$ (inset to
Fig.~\ref{T1vsT}) to Eq.~\ref{eqncho} are in good agreement with
those obtained by Cho \etal~(Fig.~\ref{Tg}) \cite{remT1analysis}.
It is remarkable that $J^\prime(x)$ is reduced by less than a
factor 2.5 between $x=0.02$, the border of the long range AF
phase, and $x=0.10$, while $c(x)$ is reduced by about a factor 10.
Inspired by the cluster spin-glass idea, Cho \etal~related the
behavior of $c(x)$ to the size of AF clusters that shrink with
doping, and $2J^\prime$ to the energy barrier for the
re-orientation of the staggered moments \cite{Cho92}. In order to
gain more physical insight into the parameters $J^\prime$ and $c$,
and into the details of spin-freezing, it would be interesting to
have theoretical predictions from different models (various forms
of cluster spin-glass, stripes, impurities, etc.).

%%%%%%%%%%%%%%%%%%%%%%%%%%%%%%%%%%%%%%%%%%%%%%%%%%%%%%%%%
\section{Summary of main results and discussion}

\subsection{Coexistence of magnetic order with
superconductivity}

Although we could study only the freezing process and not the
frozen state ($T<T_g$), our \la~$T_1$ measurements confirm
previous microscopic evidence for frozen magnetic moments below
$\sim$1~K in \lten~\cite{Niedermayer98}. In our sample, which is a
bulk superconductor according to Huh \etal~\cite{Huh00}, we
observe that the magnetic freezing is visible at all \la~sites in
the sample. This is again in agreement with Niedermayer
\etal~finding all muons probing an internal field at low $T$
\cite{Niedermayer98}. Thus, one must conclude that
superconductivity coexists with frozen magnetic
moments~\cite{rem-coexistence} (which we know to be locally
staggered~\cite{Julien99}), and this coexistence has to occur at
the microscopic scale. It is noted that the nature of the magnetic
freezing found in the superconducting phase ($x\gtrsim0.05$)
appears to be of the same kind of the cluster spin-glass freezing
reported earlier \cite{Cho92} in non-superconducting materials.

In La$_{1.88}$Sr$_{0.12}$CuO$_4$, equal magnetic and
superconducting transition temperatures were reported, based on
neutron scattering and \la~NMR lineshape
measurements~\cite{Goto94,Goto97,Suzuki98}. However, we have shown
above that the broadening of the \la~NMR line occurs well above
the freezing temperature $T_g$ (as determined by \la~NMR $T_1$ or
$\mu$SR). We suspect that the real $T_g$ in
La$_{1.88}$Sr$_{0.12}$CuO$_4$ is much lower (see
\cite{Torikai90,Kumagai94}) and thus that the onset of NMR line
broadening coincides with $T_c$ only by chance.

The coexistence found here resembles previous works in {\it e.g.}
Nd-doped \lsco~where bulk superconductivity is claimed by several
groups (see Refs. \cite{Moodenbaugh97,Ostenson97,Nachumi98} and
\cite{Kataev98} for an opposite point of view). Nevertheless, more
experimental work remains do be performed in order to fully
characterize the superconducting properties of these materials
with spin-glass like freezing. From the theoretical viewpoint, it
is clear that the cluster spin-glass freezing deserves intensive
consideration: the existence of superconductivity in such a
context and the relation to stripe physics need being addressed
more accurately (see~\cite{CSG-theory} for recent views).

Finally, it is important to note that the coexistence of frozen
moments with superconductivity does not mean that both phenomena
are somehow related or even cooperative. There is ample evidence
that they compete (see \cite{Ichikawa00} for a recent perspective
and references). In fact, the internal static field, existing at
$T\ll T_g$, is very small in \lten, certainly less than 10\% of
the value in La$_2$CuO$_4$~\cite{Niedermayer98}.

\subsection{What are the evidences for stripes in \lsco~?}

Looking at the whole body of experimental data in \lsco~with $0.06
\lesssim x\lesssim 0.12$ (Refs.
\cite{Niedermayer98,Julien99,Panagopoulos00,Julien99b,Hunt99,Kataev93,Kochelaev97,Boebinger96,Ichikawa00}
and references therein), it is now quite clear that this part of
the phase diagram shows (besides superconductivity) three
mutually-related phenomena at low temperature
($T\lesssim40-100$~K, depending on doping): charge localization
tendencies, glassy spin-freezing and Cu NMR/NQR wipeout. It is
very tempting to attribute these three features to the presence of
charge stripes. There are two principal arguments in support to
this view:

(1) Magnetic order issued from glassy spin freezing and Cu NQR
wipeout are observed in a very similar way in rare earth-doped
\lsco, where they are understood as a consequence of stripe order
\cite{Hunt99,Singer99,Suh00,Curro00,Matsumura00,Wagener97,Nachumi98,Kataev98}.
Transport properties of these materials also bear strong
similarities to those of \lsco \cite{Tajima99,Noda99,Ichikawa00}.
Of course, another important piece of argument is the identical
wave-vector for magnetic scattering in Nd-free and Nd-doped
\lsco~\cite{Kimura99,Tranquada97}.

(2) It is difficult to imagine how a magnetic state with local AF
order could exist without charge segregation at high hole-doping
level. Charge stripes represent an ideal form of such segregation.
If doped holes were randomly distributed the mean distance between
them would be about three lattice spacings in \lten. On the other
hand, the mean distance between charge stripes is of five lattice
spacings at $x=0.10$ (assuming one charge every two sites along
the stripe). The charge stripe picture clearly generates much
larger hole-poor regions (magnetic domains), and is naturally much
more favorable to spin order : charge order eliminates spin
frustration, which is otherwise large if holes are uniformly
distributed. This interpretation of the data in the "cluster
spin-glass" phase of \lsco~has repeatedly been put forward by
Emery and Kivelson~\cite{EK}.

On the other hand, {\it direct} evidence for charge stripes is
still lacking in superconducting \lsco. This is because charged
structures are evidently difficult to detect: fluctuations and
disorder can readily render diffraction methods inoperative.
However, the accumulation of indirect hints is rather
overwhelming.

\subsection{Is the wipeout fraction a measure of stripe order~?}

We have found that the Cu NQR wipeout effect could be explained by
the glassy nature of the magnetic freezing in \lten: slow spin
dynamics shorten the nuclear relaxation times $T_1$ and $T_2$ of
$^{63,65}$Cu, and these nuclei become unobservable below a
threshold value of $T_2$. Because the dynamics is spatially
inhomogeneous and the freezing occurs on a wide $T$ interval, the
NMR/NQR signal disappears only gradually on cooling. This
interpretation of the wipeout effect was also proposed by Curro
\etal~\cite{Curro00}. These authors also argue that the crossover
from Gaussian to Lorentzian $T_2$ is explained by the same
arguments. The magnetic origin of the wipeout is taken by Curro
\etal~as a strong argument against the wipeout fraction being a
measure of the "stripe order parameter". Our point of view is
somewhat less radical, although we come to the same conclusion, as
explained below.

The fact that the wipeout can be explained by glassy spin-freezing
only indicates that the relationship with stripe order, if any, is
not straightforward. Actually, this was not excluded by the
authors of Refs.~\cite{Hunt99,Singer99}. Their data in
La$_{1.6-x}$Nd$_{0.4}$Sr$_{0.09\leq x \leq 0.15}$CuO$_4$ suggest a
sharp wipeout onset, coinciding with the charge order temperature
detected by neutron scattering~\cite{Singer99}. In this sense, the
wipeout effect must be somehow related to stripe order. This is
reasonable since charge order is always followed by magnetic
order. The slowing down of spin fluctuations on approaching the
magnetic transition is in turn responsible for the wipeout effect.
However, the argument holds only because we know that there are
stripes in these materials. One should keep in mind that any other
situation with slow and inhomogeneous spin dynamics (such as
impurity doping) could produce similar wipeout.

On the other hand, it is the very nature of the wipeout effect
that makes the identification with a stripe {\it order parameter}
questionable. The wipeout fraction measures a kind of volume
fraction in which the spin dynamics is slowed down below some
threshold value, that clearly depends on experimental conditions.
This is typically an extensive quantity, while an order parameter
is usually associated to an intensive quantity like the amplitude
of a field or of a distortion. Moreover, it is even not {\it a
priori} obvious to consider the amplitude of charge order peaks in
neutron or X-ray scattering as an order parameter. If the order is
not homogeneous, the peak intensity may also reflect the volume
fraction of well-ordered stripe segments in the crystal. Finally,
we note two contradictions which need to be clarified. First, the
wipeout onset in La$_{1.65}$Eu$_{0.2}$Sr$_{0.15}$CuO$_4$ occurs at
much higher temperature for Curro \etal~($\simeq
100-150$~K~\cite{Curro00}) than for Singer \etal~($\simeq
50-70$~K~\cite{Singer99}). Second, the onset of Cu NQR wipeout
seems to occur prior to the neutron charge ordering temperature in
La$_{1.50}$Nd$_{0.4}$Sr$_{0.10}$CuO$_4$
\cite{Singer99,Ichikawa00}.

In conclusion, while the Cu NQR wipeout effect is certainly
related to stripe order in La$_{2-x-y}$Nd$_{y}$Sr$_{x}$CuO$_4$ and
in \lsco, it would not be reasonable to use it as a criterion for
the presence of charge stripes in other materials and it is not
possible to identify the wipeout fraction to a stripe {\it order
parameter}.\\

{\it Note added} - After completion of this manuscript we became
aware of a NQR work in La$_{1.48}$Nd$_{0.4}$Sr$_{0.12}$CuO$_4$
[G.B. Teitel'baum \etal, cond-mat/0007057]. A \la~wipeout effect
is found, but starting at much lower temperature than the Cu
wipeout. The authors argue that \la~$T_1$ is entirely due to
magnetic fluctuations, and they find, in agreement with
Ref.~\cite{Curro00} and our work, that slow and distributed spin
fluctuations explain the Cu NQR wipeout effect.

\section{Acknowledgments}

Thanks are due to A.H.~Moudden for cutting a piece from his
sample, to L.P.~Regnault and P.~Bordet for neutron and X-ray
characterization of the sample. We are particularly grateful to
Y.M.~Huh, J.E.~Ostenson and D.K.~Finnemore for a collaboration on
superconducting properties of \lsco~samples. We also thank
A.~Lascialfari for some SQUID measurements. Help at various stages
of this project from S.~Aldrovandi, L.~Linati, F.~Tedoldi, N.
Poulakis is gratefully acknowledged. The work in Pavia was
supported by the INFM-PRA SPIS funding. Ames Laboratory is
operated for U.S Department of Energy by Iowa State University
under Contract No. W-7405-Eng-82. The work at Ames Laboratory was
supported by the director for Energy Research, Office of Basic
Energy Sciences. The GHMFL is "Laboratoire conventionn\'e aux
universit\'es J.~Fourier et INPG Grenoble I". Support from the
FERLIN program of the European Science Foundation is
acknowledged.\\


\begin {references}
\bibitem{rem-quad} An attempt to detect a possible enhancement
of the quadrupolar relaxation channel, by comparing the recovery
of $M_z(t)$ for central and satellite NMR lines, at different
temperatures, was unsuccessful.
\bibitem{remT1value} For convenience, $T_1$ is defined as the time at which $(M_0-M_z)/M_0$
has decreased by a factor $1/e$. The values defined in this way
are close to those from a stretched exponential fit
$(M_0(t)-M_z)/M_0=\exp((-t/T_1)^\alpha)$, which we attempted here,
but found not entirely satisfactorily (this is better seen in a
logarithmic horizontal scale and linear vertical scale). However,
one should keep in mind that these values are artificially short:
a fit with the theoretical law given in the text, which contains
large numerical factors in some exponentials, leads to values
longer by about an order of magnitude.
\bibitem{remT1analysis} In order to compare with the work Cho \etal~\cite{Cho92}, $T_1$ values for
this plot where obtained from the slope of the recovery plot for
$t\rightarrow0$. For $x=0.10$, the NMR data were rescaled by a
factor of 1.55 in order to match NQR values (see
Fig.~\ref{T1vsT}).
\bibitem{rem-coexistence} Each \la~nucleus is coupled to several Cu
sites. If these Cu sites have different fluctuating frequencies,
$T_1$ is determined by the fastest relaxation channel. This means
that $T_1$ may only probe the slowest magnetic moment to which it
is coupled. So, we cannot exclude, in principle, that some
individual Cu$^{2+}$ moments do not freeze. The data show however
that "non-freezing regions" of typical size larger than a few
lattice spacings are not present in the sample. Furthermore, a
situation with very different fluctuating frequencies on
neighboring sites is quite unlikely.
\end{references}
\end{document}